\pdfoutput=1

\documentclass[hyper]{JHEP3}

\usepackage{amsmath}
\usepackage{amssymb}
\usepackage{graphicx}

\newcommand{\Eq}[1]{(\ref{#1})} %Label equation; Do \begin{equation}\label{eom} and cite Eq.~\Eq{eom}.
\newcommand{\be}{\begin{equation}}
\newcommand{\ee}{\end{equation}}
\newcommand{\bea}{\begin{eqnarray}}
\newcommand{\eea}{\end{eqnarray}}

\newcommand{\bbibitem}[1]{\bibitem{#1}} 
\newcommand{\mc}{\mathcal}
\newcommand{\mbb}{\mathbb}

\newcommand{\color}[2][]{}

\title{Multi-Field Inflation from String Theory}

\author{Per Berglund$^\spadesuit$$^\heartsuit$
and Guoqin Ren$^\spadesuit$\\
$\spadesuit$ Department of Physics, University of New Hampshire, %\\
Durham, NH 03824, USA\\
 $\heartsuit$ PH-TH Division, CERN, CH-1211 Geneva 23, Switzerland\\
E-mail:  \email{per.berglund@unh.edu},  \email{grv2@unh.edu}}

\abstract{We construct a multi-field inflationary model consisting of multiple K\"ahler moduli derived from type IIB string compactification in the large volume limit. The model consists of both heavy and light fields, with the former being frozen during the inflationary period and the latter acting as the inflaton(s). We study the evolution of all the fields during and after inflation until the preheating era when all the fields oscillate around their vacuum expectation values. Our numerical analysis shows that the curvature perturbations have an almost scale invariant power spectrum with $n_s \simeq 0.96$.}

\preprint{CERN-PH-TH/2009-237\\
UNH-09-05}

\begin{document}

\section{Introduction}
Ever since inflation was proposed in the early 1980s~\cite{guth1981}, there have been numerous inflationary models,
including scalar field slow-roll inflation~\cite{Linde:1981mu,Albrecht:1982wi}. (For a review, see~\cite{lythriotto1999, tasi}.) In recent years, there have been some promising developments in models derived or inspired by string theory, where scalar fields associated to the shape and size of the internal space, or to the positions of branes, serve as candidate inflaton fields.  Based on the work of KKLT~\cite{KKLT} and the large volume limit on moduli stabilization~\cite{vb-pb-jc-fq}\cite{vb-pb}, an interesting slow-roll inflationary model has been proposed by Conlon and 
Quevedo~\cite{0509012}, in which the inflaton is chosen to be one of the
K\"ahler moduli, which sets the size of the compactified space. 
(For some more recent work along these lines, see \cite{08043653} and \cite{0906.3711}.)
Because the moduli only appear exponentially in the scalar potential, the potential is very flat along the directions of the light fields, which is ideal for obtaining successful inflation and a graceful exit to the normal Friedmann-Robertson-Walker (FRW) universe. 

In this type of scenario, due to the presence of more than one modulus, we should expect to see a multi-field inflationary scenario in which all the fields play a role. 
In addition, due to the nature of cosmological models arising from string compactifications, the fields are not canonically normalized, i.e., the metric is in general neither diagonal nor field independent, and the scalar potential is highly non-trivial.
This typical feature of string theory, leads to a highly coupled dynamical situation, in which, in principle, the motion of any one field impacts the evolution of the other fields.
Therefore, we need to understand how each modulus and its perturbations evolve during and after inflation. In particular, we want to understand the difference between the ``heavy'' moduli and the ``light'' moduli. (For some earlier work, addressing the issue of the ``very high energy physics" in inflation, see~\cite{Burgess:2002ub}.) To accomplish this, it is necessary to explicitly solve the equations of motion for all the moduli and their perturbations. We study the resulting power spectra and the spectral indices of the perturbations, and
compare the result with the usual single field scenario. 

In \cite{0509012}  the general idea of K\"ahler moduli inflation was outlined and  a single modulus inflationary case was demonstrated by assuming that all other moduli are already in the final vacuum states. The same assumption was  also made in \cite{08043653} with two light fields.  We will show that this assumption is redundant and may be dropped when all the moduli are taken into account. This is consistent with the numerical analysis in \cite{0906.3711} which further pointed out that there exists a large region of parameter space within which the inflationary solutions fit the observations.  
Our work considers the general case of both multiple heavy and light fields. Due to the multi-field nature of our inflationary scenario, i.e., the existence of several light fields, we also compute the isocurvature perturbations. 

The outline of the paper is as follows.
We first solve the perturbation equations of the scalar fields in Section \ref{sfp}. We also demonstrate the difference between the perturbation solutions of the heavy and light fields.  In Section \ref{tcip}, we study the curvature and isocurvature perturbations. In Section \ref{tpkms}, we  introduce the moduli stabilization in type IIB string theory and the effective potential.  In the numerical analysis, in Section \ref{ms}, we investigate a few examples of the multi-field inflationary model and compute the spectra and tilts (spectral indices) which are used to make contact with observations. We conclude and summarize our results in Section \ref{conc}. The detailed analysis of the Hankel functions, needed to study the effects of the heavy fields, can be found in Appendix A.

\section{Scalar Field Perturbations}
\label{sfp}
In this section, we study the scalar perturbations in a general multi-field model, following previous work, in particular that by Byrnes and Wands~\cite{0605679}. We explicitly show how the perturbations for the heavy fields are suppressed.
\subsection{The Background Equations of Motion}
Let us consider the scalar-tensor field theory of gravity, with the action of the standard form
\begin{equation}\label{action}
S = \int d^4 x \sqrt{-g}[ \frac{R}{2 \kappa^2} + \frac{1}{2}h_{ab} g^{\mu\nu} \partial_\mu\phi^a\partial_\nu\phi^b -V(\phi)]
\end{equation}
where $h_{ab}$ is the metric on the space of fields, $\phi^a$-space. From now on we will work in units where $\kappa^2 =8\pi G_N ={M_p}^{-2} = 1$\footnote{$M_p$ is the Planck mass, $\sim 2.4\times 10^{18}$ Gev.}. 
\\

The background spacetime metric $g_{\mu\nu}$ is chosen to be the Friedmann-Robertson-Walker (FRW) metric. The line element  is given by
\begin{equation}
ds^2 = -dt^2 + a^2(t)\gamma_{ij} dx^i dx^j = a^2(\eta)(-d\eta^2+\gamma_{ij} dx^i dx^j)
\end{equation}
where
\begin{equation}
\gamma_{ij} = \delta_{ij} [1+ 4kr^2]^{-2}\,.
\end{equation}
We will work in the flat universe in which $k=0$
\begin{equation}\label{flatspace}
\left( \begin{array}{cccc}
-1 & 0&0&0 \\
0 & a^2(t) & 0&0 \\
0 & 0& a^2(t)&0 \\
0 &  0&0 & a^2(t)
\end{array} \right ).
\end{equation}

The background fields, $g_{\mu\nu}$ and $\phi^a$, satisfy the usual equations of motion obtained by varying
the action $S$~\Eq{action}. This gives rise to the following field equations for the background fields $\phi^a$,
\begin{equation}\label{eom}
[\frac{D}{dx^\mu}+ \frac{1}{\sqrt{-g}}(\partial_\mu \sqrt{-g})]h_{ab}g^{\mu\nu} \partial_\nu \phi^b + V_{,a} = 0\,,
\end{equation}
where $\frac{D}{dx^\mu}$ is the covariant derivative. For flat space,  $g = -a^6$, and assuming that the fields are homogeneous, $\phi^a=\phi^a(t)$, Eq.~\Eq{eom} becomes 
\begin{equation}
\frac{d^2\phi^a}{dt^2} + \gamma^a_{bc}\frac{d\phi^b}{dt}\frac{d\phi^c}{dt} + 3H \frac{d\phi^a}{dt} + h^{ab}\frac{\partial V}{\partial \phi^b} =0\,,
\end{equation}
where $\gamma^a_{bc}$ is the Christoffel connection on the space of fields.
By introducing the notation
$\dot{\phi^a} = \frac{d\phi^b}{dt}$, $\ddot{\phi^a}= \frac{D}{dt} \dot{\phi^a} = \dot{\phi^b}\nabla_b \phi^a $,\footnote{$\frac{D}{dt}$ denotes the covariant derivative on the coordinate space, while $\nabla_b$ is the covariant derivative on the field space (labeled by $b$).} the field equations for $\phi^a$ can be further simplified
\begin{equation}
\ddot{\phi^a} + 3H \dot{\phi^a} + V^{,a} =0\,.
\end{equation}
Varying $S$ with respect to the spacetime metric gives the Einstein equations
\begin{align}\label{ee}
3 H^2 = &\frac{1}{2} \dot{\phi^a}\dot{\phi_a} + V\,,\\
\dot{H} = & -\frac{1}{2} \dot{\phi^a}\dot{\phi_a}\,.
\end{align}

The background equations determine the evolution of the inflaton and how long inflation will last, the e-folding time, $N$. In most cases, the light fields and the heavy fields\footnote{The ``light'' fields have effective masses less than the Hubble parameter, $m_h^2 = V_{II} < V \sim H^2$, while the ``heavy'' fields are heavier than the Hubble parameter.} are decoupled, i.e., the heavy fields remain constant most of the time when the light fields is rolling. Thus, the inflaton is always one, or a subset, of the light fields which controls the dynamics of inflation. However, in what follows, we will not make any assumptions about which fields are frozen and which are dynamic, but rather establish this fact as part of the calculation.

\subsection{Perturbations}

The most general scalar 
perturbations on the background spacetime metric is~\cite{0605679}~\cite{mukhanov}
\begin{equation}\label{psm}
ds^2 = -(1+2A)dt^2 + 2 a B_{;i}dx^i dt + a^2[(1-2 \psi)+ 2E_{;ij}]dx^i dx^j
\end{equation}
The associated equation of motion for the perturbed fields are~\cite{nast}
\begin{equation}\label{peom}
\frac{D^2}{dt^2}\delta \phi^a + 3H \frac{D}{dt} \delta \phi^a + R^a_{cbd} \dot{\phi^c}\dot{\phi^d}\dot{\phi^b} + \frac{k^2}{a^2} \delta\phi^a + \delta \phi_b V^{;ab} = \frac{\delta \phi_b}{a^3} \frac{D}{dt}[\frac{a^3}{H} \dot{\phi^a}\dot{\phi^b}]
\end{equation}
where the covariant derivatives and the Riemann curvature tensor are all evaluated on the field space. Here the fields, $\phi^a$ and their perturbations, $\delta \phi^a$, are evaluated at a particular  (comoving) wavenumber $k$, i.e.,
\bea
\phi^a &=& \phi^a_k (t) = \int d^3x \phi^a(\boldsymbol{x},t) e^{ikx} \\
\phi^a(\boldsymbol{x},t) &=& \frac{1}{(2\pi)^3} \int d^3k \phi^a_k(t) e^{-ikx}\,,
\eea
and similar for the perturbations.
The perturbed Einstein equations are 
\bea%\begin{equation}
\label{pee1}
- 6H^2A - 2k^2 H \dot{E} &=& - A \dot{\phi^a}\dot{\phi_a} + \dot{\phi^a}\frac{D}{dt}\delta \dot{\phi_a}\\
%\end{equation}
%and 
%\begin{equation}
\label{pee2}
2 HA &=& \dot{\phi_a} \delta \phi^a\,.
\eea%\end{equation}\\

To simplify the problem of solving the equations of motion for the perturbations, we work with the canonical field-space metric\footnote{In general, the moduli space metric, $h_{ab}$ is neither canonically normalized nor field independent. However, we show that in the class of models we are considering it is possible to make a field-dependent field redefinition such that the metric remains (approximately) flat throughout and after the inflationary period.} 
in the spatially flat gauge. Eq.~\Eq{peom} then becomes~\cite{0605679}
\begin{equation}\label{cpeom}
\ddot{\delta \phi^I} + 3 H \dot{\delta \phi^I} + \frac{k^2}{a^2}\delta \phi^I + \sum_J[V^I_J - \frac{8 \pi G}{a^3} \frac{D}{dt} (\frac{a^3}{H} \dot{\phi^I}\dot{\phi_J})] \delta\phi^J = 0\,.
\end{equation}
To solve this equation, we use the conformal time $\tau {=}\int a^{-1} dt$ instead of the cosmic time $t$:
\begin{align}
&\frac{D}{dt} = \frac{1}{a} \frac{D}{d\tau}\nonumber \\
&\frac{D^2}{dt^2} = \frac{1}{a} \frac{D}{d\tau}(\frac{1}{a} \frac{D}{d\tau}) = -\frac{\dot{a}}{a^2}\frac{D}{d\tau} + \frac{1}{a^2}\frac{D^2}{d\tau^2}\,,
\end{align}
where $'$ denotes differentiation with respect to $\tau$ and $\dot{}$ denotes differentiation with respect to $t$.
Then, by making the change of variables, $\delta \phi^I = \frac{1}{a} u_I$, where the superscripts get changed to subscripts for later convenience, we have
\begin{align}
\dot{\delta \phi^I} &= \frac{\dot{u_I}}{a} - \frac{u_I\dot{a}}{a^2} = \frac{{u_I}'}{a^2} - \frac{u_I\dot{a}}{a^2} \nonumber \\
\ddot{\delta\phi^I} & = \frac{\ddot{u_I}}{a} - 2 \frac{\dot{u_I}\dot{a}}{a^2} + 2\frac{u_I {\dot{a}}^2}{a^3} - \frac{u_I \ddot{a}}{a^2} = \frac{{u_I}''}{a^3} - 3 \frac{\dot{a}}{a^3}{u_I}' + (2\frac{\dot{a}^2}{a^3} - \frac{\ddot{a}}{a^2})u_I\,.
\end{align}
We also use the slow-roll approximation during inflation,
\begin{equation}
\epsilon= \frac{\frac{1}{2} \dot{\phi^I}\dot{\phi_I}}{H^2} < <1
\end{equation}
and integrate the conformal time by parts~\cite{astro-ph/0106020}\cite{0908.0585},
\begin{equation}\label{tau}
\tau \simeq -\frac{1+\epsilon}{aH} +  O(\epsilon^2)\,.
\end{equation}

Thus, in terms of the conformal time~\Eq{tau},  Eq.~\Eq{cpeom} becomes
\begin{equation}\label{ueq}
u_I'' + (k^2 - \frac{2}{\tau^2}) u_I = \frac{3}{\tau^2}\sum_J M_{IJ} u_J
\end{equation}
where %we have used the approximation~\Eq{tau}, and 
the matrix $M_{IJ}$ is given by
\begin{equation}\label{2drm}
M_{IJ} = \epsilon \delta_{IJ} + 2 \epsilon_{IJ} - \eta_{IJ} -\frac{2}{3}\epsilon_{IJ}(\frac{\ddot{\phi^I}}{\dot{\phi^I}}+\frac{\ddot{\phi^J}}{\dot{\phi^J}}) + \frac{2}{3}\epsilon \epsilon_{IJ} + O(\epsilon^2)
\end{equation}
and the multi-field slow-roll parameters are defined as follows
\begin{align}
&\epsilon_{IJ} = \frac{1}{2} \frac{\dot{\phi}^I\dot{\phi}^J}{H^2} \\
&\eta_{IJ} = \frac{1}{V}\frac{\partial^2V}{\partial \phi^I \partial \phi^J} =\frac{V_{IJ}}{V} \nonumber\,.
\end{align}

Note that the last term, $\frac{2}{3}\epsilon \epsilon_{IJ}$,  is second order in slow-roll parameters and may be ignored. The forth term in~\Eq{2drm} can also be treated as a second order term  for the light fields.   For the heavy fields, this term may be a first order term, $\sim \mathcal{O}(1)\epsilon_{IJ}$. For completeness sake, we will keep all the terms in our analysis throughout this paper.

To solve~\Eq{ueq}, we have to decouple the equations by making a rotation $U$ such that
\begin{equation}
U^{-1}MU = diag\{\lambda_I\}\,,
\end{equation}
where $\lambda_I$ are the eigenvalues of $M$. Then $U$ is given by the similarity transformation 
\begin{equation}
U= \left(\begin{array}{c}
\vec{g_1}\text{ }\vec{g_2}\text{ }...\text{ }\vec{g_n}
\end{array}\right)\,,
\end{equation}
where $\vec{g_i}$ are the eigenvectors of of $M$.

Thus, by introducing the new fields, $v_I$,
\begin{equation}
\label{t_function_1}
u_I = U_{IJ} v_J\,,\quad{\rm or} \quad
%\label{t_function_2}
v_I = U^{-1}_{IJ}u_J\,,
\end{equation}
we get that Eq.~\Eq{ueq} is decoupled
\begin{equation}\label{decoupled_perturbation_equations}
v_I'' + (k^2 - \frac{\mu_I^2 - \frac{1}{4}}{\tau^2}) v_I = 0\,,
\end{equation}
where 
\begin{equation}\label{coeff_bessel}
{\mu_I}^2 = \frac{9}{4} + 3\lambda_I\,.
\end{equation}

Now we want a solution satisfying the Minkowski-like vacuum initial conditions~\cite{tasi} when $k \tau_i \gg 1$ ($k\gg aH$)
\begin{equation}\label{min_vac_ini_cond}
v_I \sim \frac{e^{-ik\tau_i}}{\sqrt{2k}}\,.
\end{equation}
The solution is
\begin{equation}\label{solpem}
v_I = \frac{\sqrt{\pi}}{2} e^{i\frac{(\mu_I + \frac{1}{2})\pi}{2}} (-\tau)^{\frac{1}{2}} H^{(1)}_{\mu_I}(-k\tau) \hat{e}_I(k)\,,
\end{equation}
where $\hat{e}_I$  are the normalized  Gaussian variables\footnote{The fluctuation can be treated as a random field which is a \emph{Gaussian process}. The homogeneous universe can be divided into a set of sample space with different values of \emph{random fields} mapped on it.}, satisfying~\cite{0605679}\cite{mukhanov-book}\cite{07040212}
\begin{align}
&\langle \hat{e}_I(k)\rangle = 0 \\
& \langle \hat{e}_I(k)\hat{e}_J^*(k')\rangle = \delta_{IJ} \delta^3(k-k')\,.
\end{align}

\subsection{The Asymptotic Solution}
We are mainly interested in the solution after Hubble exit when $k < aH$ or $k\tau \to 0$.
\paragraph{For small $\lambda_I$,}\hskip-0.3cm
\footnote{These solutions are related to the light fields. Recalling~\Eq{2drm}, all the components in $M_{IJ}$ related to the light fields are first order in slow-roll parameter. Thus the corresponding eigenvalues $\lambda_I$ are small, too.}
\[
\mu_I = \frac{3}{2} + \lambda_I + O({\lambda_I}^2), \text{ for } \lambda_I \ll 1
\]
Eq.~\Eq{solpem} becomes
\begin{equation}\label{solpem0}
v_I \simeq ie^{i\frac{\pi}{2}\lambda_I}(1 + C\lambda_I)\frac{1}{\sqrt{2k}}(-k\tau)^{-1-\lambda_I}\hat{e}_I(k), \text{ for } k\tau \to 0\,,
\end{equation}
where $C = 2 -log2 -\gamma$ ($\gamma$ is the Euler-Mascheroni constant).

\paragraph{For large and negative $\lambda_I$,}\hskip-0.4cm
\footnote{ These solutions are related to the heavy fields. For the heavy fields, the relevant components in $M_{IJ}$(and thus $\lambda_I$) are dominated by the diagonal elements of the matrix ($M_{II}$) which are the curvature of the potential, $\sim - \frac{V_{II}}{V}\propto -\frac{m_I^2}{H^2}$.} the order $\mu_I$ is complex (with a large imaginary part) and $|\mu_I| >>1$.  We need to expand the Hankel function of large complex order. (For more details, see Appendix A). 

Recall~\Eq{coeff_bessel}, when $\lambda_I$ is large and negative,
\begin{equation}\label{coeff_bessel_expan}
\mu_I = \sqrt{\frac{9}{4} + 3\lambda_I} \approx \pm i\rho_I, \text{ } \rho_I >0, \text{ } |\rho_I| > >1\,,
\end{equation}
which is (almost) purely imaginary. 
Using~\Eq{Hankel_approx_l_im_ord_s_arg}, the solution~\Eq{solpem} becomes
\begin{equation}\label{solution_perturbation_approx_l_im_ord_s_arg}
v_I \simeq \frac{1}{\sqrt{2}} (-\tau)^{\frac{1}{2}} (1+\rho_I^2)^{-\frac{1}{4}} e^{1 +(\alpha-\frac{\pi}{2}) \rho_I} e^{i\frac{\pi}\,,{4}}\omega\hat{e}_I(k)
\end{equation}
where \footnote{The dimension of~\Eq{solution_perturbation_approx_l_im_ord_s_arg} is $|\tau|^{\frac{1}{2}}$, while the dimension of~\Eq{solpem0} is $|k|^{-\frac{1}{2}}$. They are the same since $k\tau$ is dimensionless.}
\begin{equation}\label{dots}
\omega =e^{-\rho \pi} (\frac{z}{2})^{-i\rho} e^{i(\frac{\alpha}{2} -\rho + \rho log\sqrt{1 + \rho^2})} - (\frac{z}{2})^{i\rho}e^{-i(\frac{\alpha}{2} -\rho + \rho log\sqrt{1 + \rho^2})}
\end{equation}
and $z=-k\tau$ .

Note that $|\omega| \neq 1$ in general. But for large $\rho$,
\begin{equation}\label{exp_supp}
|\omega|^2 = 1 + 2 cos\beta(z) e^{-\rho \pi} + e^{-2\rho \pi} \approx 1
\end{equation}
where $\beta(z)$ is some function of $z=-k\tau$ according to~\Eq{dots}. As we can see, the dependence on $k$ for $|\omega|^2$ (and hence the power spectrum), mainly given by the second term in~\Eq{exp_supp},  is exponentially suppressed. Hence, in the limit $\rho_I \to \infty$
\begin{equation}\label{solution_perturbation_approx_l_im_ord_s_arg_large}
v_I \propto (-\tau)^{\frac{1}{2}} (1+\rho^2)^{-\frac{1}{4}} e^{1 +(\alpha-\frac{\pi}{2}) \rho_I} \overset{\rho_I \to \infty}{\longrightarrow} (-\tau)^{\frac{1}{2}} \frac{e}{\sqrt{\rho_I}} \,,
\end{equation}
and the solution  for large $\rho$ is suppressed by a factor of $\frac{1}{\sqrt{\rho}}\sim \frac{1}{\sqrt{m}}$.

This asymptotic solution can also be partially obtained from the following consideration. Consider the perturbation equation~\Eq{decoupled_perturbation_equations} when $\mu_I^2$ is large and negative and $k\tau$ is small. It is approximate to the equation
\[
v_I'' + (- \frac{\mu_I^2}{z^2}) v_I = 0\,,
\]
where we change the variable $\tau \to z = -k\tau$. The solution of this equation is
\[
v_I(z) \propto  z^{\frac{1}{2}} z^{\pm i \sqrt{-\frac{1}{4}-\mu_I^2}}\,,
\]
which behaves similarly as~\Eq{solution_perturbation_approx_l_im_ord_s_arg}.

In summary, the perturbation solutions are
\[
v_J(-k\tau)\overset{k\tau \to 0}{ \simeq }\left\{ 
\begin{array}{l l}
ie^{i\frac{\pi}{2}\lambda_J}(1 + C\lambda_J)\frac{1}{\sqrt{2k}}(-k\tau)^{-1-\lambda_J}e_J(k),& \quad |\lambda_J| <<1\,, \\
\frac{1}{\sqrt{2}} (-\tau)^{\frac{1}{2}} (1+\rho_J^2)^{-\frac{1}{4}} e^{1 +(\alpha-\frac{\pi}{2}) \rho_J} e^{i\frac{\pi}{4}}\omega e_J(k), & \quad -\lambda_J >1\,,
\end{array} \right.
\]
where $\rho_J = \sqrt{-(\frac{9}{4}+3 \lambda_J)}$\,. 

\section{The Curvature and Isocurvature Perturbations}
\label{tcip}
\subsection{The Perturbations and Power Spectra}
It is convenient to decompose the scalar field perturbations into \emph{adiabatic perturbation}\footnote{Also called curvature perturbations.}(parallel to the background trajectory) and \emph{entropy perturbation}\footnote{Also called non-adiabatic perturbations or isocurvature perturbations.}(orthogonal to the background trajectory). We then define the adiabatic component
\begin{equation}\label{sigs}
\delta \sigma = \sum_I \delta \phi_I cos \theta^I  
\end{equation}
and the non-adiabatic component
\begin{equation}
\delta s^2 = \sum_I {\delta \phi_I}^2 - \delta \sigma^2\,,
\end{equation}
with
\be\label{cos-theta}
cos\,\theta^I = \frac{\dot{\phi_I}}{\sqrt{\sum_I \dot{\phi_I}^2}} =\frac{\dot{\phi_I}}{\dot{\sigma}}
\ee
and
\be\label{pert-light}
\delta\phi^I = \frac{1}{a}U_{IJ}v_J\,.
\ee
Note that in~\Eq{pert-light}, the sum over the $v_J$ is for the light solutions only, since the perturbations of the heavy fields are strongly suppressed (see the discussion in Section \ref{sec:evo_after_hubble} for more details).

By definition, the two-point correlation functions (i.e., power spectra) are given by
\begin{equation}
C_{xy}\delta^3(\vec{k}-\vec{k}') = \frac{4\pi k^3}{(2\pi)^3} \langle x(\vec{k})y^*(\vec{k'})\rangle
\end{equation}
where $x,y$ = $\delta\sigma, \delta s$. 
For example, 
\begin{align}\label{apsd}
C_{\sigma\sigma}(k)\delta^3(k-k') &= \frac{k^3}{2\pi^2}\left<\delta\sigma^*\delta\sigma\right> \nonumber \\
& =\frac{k^3}{2\pi^2}  \sum_{IJ}\left<\delta \phi_I^* \delta \phi_J\right>   cos \theta^I cos \theta^J \nonumber \\
&=\frac{k^3}{2\pi^2a^2}\sum_{IJp}cos \theta^I cos \theta^JU_{Ip}U_{Jp}\left<v_p^* v_p\right>  
\end{align}
where we used
\[
\delta \phi_I = \frac{1}{a} U_{IJ} v_J
\]
For $v_J$ in the above expression, we have to omit the heavy solutions (where $-\lambda_J >1$) since they decay rapidly due to the expand of the universe (see discussions in the next section). 

We next turn to the calculation of the power spectra using these correlation functions. 

\subsubsection{Curvature Perturbations}
The comoving\footnote{Comoving means absent of peculiar motion. Comoving observers, such as large galaxies and galaxy clusters,  measure zero momentum density at their own positions~\cite{lythriotto1999}. Their position, $\vec{x}$, is time-independent in the unperturbed universe. Their physical coordinate is $ a(t) \vec{x}$.} curvature perturbation is defined by
\begin{equation}
\mathcal{R} = \psi + \frac{H}{\dot{\sigma}} \delta\sigma
\end{equation}
In spatially flat gauge ($\psi =0$), the curvature perturbations  
become~\cite{tasi}\cite{0605679}
\begin{equation}
\mathcal{R} = \frac{H}{\dot{\sigma}} \delta\sigma\,.
\end{equation}

In multi-field inflation, in addition to the curvature perturbation, the isocurvature perturbations arise from the fluctuations orthogonal to the background trajectory
\begin{equation}
\mathcal{S}= \frac{H}{\dot{\sigma}} \delta s\,.
\end{equation}

The power spectrum of $\mathcal{R}$ is defined as the expectation value of the Fourier components, which is just the ensemble average of the perturbations
%\begin{equation}
\bea
\left<\mathcal{R}_k\mathcal{R}_{k'}\right>_* &=&\frac{2\pi^2}{k^3}P_\mathcal{R}(k)\delta(\mathbf{k}-\mathbf{k'})|_*\\
%\end{equation}
%\begin{equation}
\label{cps}
P_\mathcal{R}(k)_*&=& \frac{H^2}{\dot{\sigma}^2}C_{\sigma\sigma}(k)|_*\,.
\eea
%\end{equation}
Because of slow-roll approximation, the spectrum is usually calculated  at Hubble crossing, denoted by $_*$. In practice, Hubble crossing is often taken to be 50 or 60 e-foldings before the end of inflation\cite{arXiv:astro-ph/9607038, arXiv:0803.0547}.  Due to the presence of isocurvature perturbation, the spectrum can change after Hubble crossing, which will be discussed in the following section. 

The power spectrum can be expanded around some $k_0$~\cite{08121079} \cite{peiris}
\begin{equation}
P_\mathcal{R}(k) = P_\mathcal{R}(k_0) (\frac{k}{k_0})^{n_s(k_0) - 1 + \frac{1}{2}\alpha ln\frac{k}{k_0}}\,,
\end{equation}
where 
\begin{equation}
n_s(k) -1 = \frac{d ln P_\mathcal{R}^2(k)}{d lnk}, \text{ } \tilde{\alpha} = \frac{d n_s}{d lnk}\,.
\end{equation}
We have assumed that the momentum dependence of the running, $\tilde{\alpha}$, can be neglected. In addition, $\tilde{\alpha}$ itself is of second order in slow-roll and should be small.
We next turn to the power spectrum of the isocurvature fluctuation, $P_\mathcal{S}$, and the correlation power spectrum, $P_\mathcal{RS}$. 

\subsubsection{The Isocurvature Perturbations}
\setcounter{equation}{0}  % reset counter 
The power spectrum of the isocurvature fluctuation, $P_\mathcal{S}$, and the correlation power spectrum, $P_\mathcal{RS}$, can be obtained in a similar way to the curvature perturbations. 
The non-adiabatic component has the general form
\begin{equation}\label{dsiso}
\delta s = \delta s_l =\beta^{I}\delta\phi_{I}
\end{equation}
For example, in a four-field model containing two heavy fields $(\phi_1, \phi_2)$ and two light fields $(\phi_3, \phi_4)$,
\[\beta^{I}=(1, 1, -cos \theta^{4}, cos \theta^{3})  \]
where 
\[cos\theta^{I} =\frac{\dot{\phi}_{I}}{\dot{\sigma}} \]
In a three field model, where there are two heavy fields $(\phi_1, \phi_2)$ and one light fields $\phi_3$,
\[\beta^{I}= (1, 1, 0)  \]

For the decoupled case, as will be discussed in section~\ref{sec:evo_after_hubble}, we can totally ignore the heavy fields, and the coefficients reduce to the simpler forms
\[\beta^{I}\simeq \begin{cases}(0, 0, -cos \theta^{4}, cos \theta^{3}) \\
(0, 0, 0)
\end{cases} \] 
for the four- and three-field models respectively. 

The perturbations are then given by
\begin{equation}\label{vtophi}
\delta \phi_{I} = \frac{1}{a} U_{IJ} v_J, \quad \text{ summed over the light } v_J\text{'s}\,,
\end{equation}
from which the correlation functions follow
\bea%\begin{equation}
\label{sscor}
\left<\delta s^*\delta s\right> &=& \left<\beta^I\delta\phi_I^* \beta^J\delta\phi_J\right> =  \frac{1}{a^2}\beta^I\beta^J U_{Ip}U_{Jp} \left<v_p^* v_p\right>\\
%\end{equation}
%\begin{equation}
\label{sigscor}
\left<\delta \sigma^*\delta s\right>&=& \left<cos\theta^I\delta\phi_I^* \beta^J\delta\phi_J\right> =  \frac{1}{a^2}cos\theta^I\beta^J U_{Ip}U_{Jp} \left<v_p^* v_p\right>\,.
\eea%\end{equation}
As before, the heavy $v_p$'s are ignored in the calculation. 

Thus, the power spectrum of the isocurvature fluctuation, $P_\mathcal{S}$, and the correlation power spectrum, $P_\mathcal{RS}$, at Hubble crossing are given by
\bea%\begin{equation}
\label{isocps}
P_\mathcal{S}(k)_*&=& \frac{H^2}{\dot{\sigma}^2}C_{ss}(k)|_*\\
%\end{equation}
%\begin{equation}
\label{crocps}
C_\mathcal{RS}(k)_*&=& \frac{H^2}{\dot{\sigma}^2}C_{\sigma s}(k)|_*
\eea%\end{equation}
where the two-point functions are given by
%\begin{equation}
\bea
C_{ss}(k)\delta^3(k-k') &=& \frac{k^3}{2\pi^2}\left<\delta s^*\delta s\right>=\frac{k^3}{2\pi^2a^2}\beta^I\beta^J U_{Ip}U_{Jp} \left<v_p^* v_p\right>\\
%\end{equation}
%\begin{equation}
C_{\sigma s}(k)\delta^3(k-k') &=& \frac{k^3}{2\pi^2}\left<\delta \sigma^*\delta s\right>=\frac{k^3}{2\pi^2a^2}cos\theta^I\beta^J U_{Ip}U_{Jp} \left<v_p^* v_p\right>\,.
%\end{equation}
\eea
For future reference, it is convenient to define a dimensionless measure of the correlation angle between the power spectra~\cite{0605679},
\begin{equation}\label{corang}
cos\,\Delta = \frac{C_{\mathcal{R}\mathcal{S}}}{{P_{\mathcal{R}}}^{\frac{1}{2}}{P_{\mathcal{S}}}^{\frac{1}{2}}}\,.
\end{equation}

\subsection{The Evolution of Perturbations After Hubble Exit}
\label{sec:evo_after_hubble}
For purely adiabatic perturbations, the curvature perturbation is a constant on super-horizon scales during the primordial era\footnote{The primordial era is defined as the period between Hubble exit and Hubble entry when the comoving scale, equals the Hubble scale, $\frac{a}{k} = \frac{1}{H}$.}\cite{lythriotto1999}\cite{tasi}. In this case, the observable perturbations are calculated at horizon crossing.
However, as Wands \emph{et al}. have pointed out~\cite{0605679}\cite{0003278}\cite{0205253}, the presence of entropy perturbations can change the curvature perturbation. In general, the time dependence of the curvature and isocurvature perturbation has the following form~\cite{0205253}\cite{0107502}
\begin{equation}\label{timedepcur}
\dot{\mathcal{R}} = \alpha H \mathcal{S} 
\end{equation}
\begin{equation}\label{timedepiso}
\dot{\mathcal{S}} = \beta H \mathcal{S} 
\end{equation}
or in terms of the transfer functions
\begin{equation}\label{tranmatrix}
\begin{pmatrix}
R \\
S
\end{pmatrix}=
\begin{pmatrix}
1 & T_{\mathcal{R}{S}} \\
0 & T_{\mathcal{S}{S}}
\end{pmatrix}
\begin{pmatrix}
R_* \\
S_*
\end{pmatrix}
\end{equation}

The curvature perturbation on super-horizon scales is conserved if the perturbations are purely adiabatic or if the non-adiabatic perturbation is negligible. This general conclusion does not even depend on the slow-roll approximation or the form of the gravitational field equations (the specific theory of gravity)~\cite{0003278}.  

As we can see from the solutions of the perturbation equations~\Eq{solpem0} and~\Eq{solution_perturbation_approx_l_im_ord_s_arg}, for each scale $(1/k)$, the spectrum of the perturbations with $-\lambda_J >1$ decay rapidly as the universe expands, $\left<\frac{1}{a^2}|v |^2\right> \sim \frac{1}{a^3}$. The spectrum of the perturbations with $|\lambda_J| <<1$, on the other hand, changes slowly,  
$\left<\frac{1}{a^2}|v |^2\right> \sim \left<\frac{1}{a^2}|v |^2\right>_*[1 + O(\epsilon) + O(\frac{m_l}{H})]$,
to leading order in the slow-roll parameters and the masses of the light fields over Hubble parameter. Thus we can ignore the contributions from the former and simplify the calculation.

Recall~\Eq{t_function_1} or
\[
\delta \phi_I = \frac{1}{a} U_{IJ} v_J
\]
where $U_{IJ}$ is the transfer matrix determined by the mass matrix $M$ of~\Eq{2drm}. If we assume that the heavy fields and the light fields are decoupled in such a way that the cross components $M_{IJ}$ (or $M_{JI}$), with $I$ and $J$ identified as light fields and heavy fields respectively, are subdominant compare to the non-cross components, then 
\begin{equation}
\delta \phi_{I_l}  \approx \frac{1}{a} U_{I_lJ_l} v_{J_l}, \quad I_l \& J_l \text{ denote the light fields},
\end{equation}
and
\begin{equation}
\delta \phi_{I_h}  \approx \frac{1}{a} U_{I_hJ_h} v_{J_h}, \quad I_h \& J_h \text{ denote the heavy fields}.
\end{equation}
This is true for most inflationary models encountered so far. For counterexamples, one has to use the full transfer matrix as in~\Eq{t_function_1}. Under the above assumption, the perturbations of the light fields,$\left<|\delta\phi_{I_l}|^2\right> \sim \left<|\delta\phi_{I_l}|^2\right>_*[1 + O(\epsilon) + O(\frac{m_l}{H})]$, decay much slower than the perturbations of the heavy fields, $\left<\delta\phi_{I_h}|^2\right> \sim \frac{1}{a^3}$. Therefore, in this case, one can neglect the contributions from the heavy fields when we calculate the curvature and isocurvature perturbations since the fluctuations in the heavy fields are strongly suppressed\footnote{We can always do this unless the amplitude of the non-adiabatic fluctuation is greatly amplified at the end of inflation in the preheating stage\cite{0003278}\cite{9704452}\cite{08103913}.}.

If there is a single light field (with all other fields being heavy), the perturbations are purely adiabatic and the comoving curvature perturbation remains constant during inflation. If there is more than one light field, the cosmological inflation is driven by all the light fields. In addition to the adiabatic perturbation, they also produce entropy perturbation orthogonal to the background trajectory. In this case, the curvature perturbation is no longer a constant on super-horizon scales during inflation. The coupling between the entropy perturbation and the adiabatic perturbation, given by the~\Eq{timedepcur} and~\Eq{timedepiso}, determines the evolution of the perturbations during and after inflation.

In a typical two light field inflationary model, for example, with arbitrary potential and arbitrary background trajectory, it was shown~\cite{0605679}\cite{0205253} that the scale-dependence of the observable spectra is determined by the slow-roll parameters at Hubble exit and the current observable cross-correlation.  The amplitude of the power spectra are determined by the power spectra calculated at Hubble exit and the transfer functions which parameterize the detailed physics after Hubble exit until the end of reheating, given by~\Eq{tranmatrix}~\cite{0205253}
\begin{align}\label{obspectra}
P_\mathcal{R} &= (1 + T_{\mathcal{R}\mathcal{S}}^2)  {P_{\mathcal{R}}}_* + 2T_{\mathcal{R}\mathcal{S}}{C_{\mathcal{R}\mathcal{S}}}_* \nonumber \\
P_\mathcal{S} & = T_{\mathcal{S}\mathcal{S}}^2{P_{\mathcal{S}}}_*\nonumber \\
C_{\mathcal{R}\mathcal{S}} & = T_{\mathcal{S}\mathcal{S}}{C_{\mathcal{R}\mathcal{S}}}_*+ T_{\mathcal{R}\mathcal{S}}T_{\mathcal{S}\mathcal{S}}{P_{\mathcal{S}}}_*
\end{align}

\section{The Potential and K\"ahler Moduli Stabilization}
\label{tpkms}
In what follows we focus on a particularly inflationary model derived from string theory consisting of multiple K\"ahler moduli, in the large volume limit (also known as the Large Volume Scenario)~\cite{vb-pb-jc-fq}\cite{CQS}.  We adopt the model originally proposed by Conlon and Quevedo in \cite{0509012} and subsequently studied in~\cite{08043653}\cite{0906.3711}. For more details, and in particular the conventions, see~\cite{CQS}.

Supergravity in a four dimensional theory with ${\cal N}=1$ supersymmetry is completely specified by a K\"ahler potential $\mc{K}$ and superpotential $W$. 
The K\"{a}hler potential is a real function of the complex scalar fields, while the superpotential is a holomorphic function depending only on the $\phi^i$, and not the complex conjugate, $\bar{\phi}^i$.
Focusing on the  dynamics of the scalar fields relevant for inflation, the supergravity action is (we will work in the Einstein frame, and in units where $M_P^2=1$)
\begin{equation}
S_{{\cal N}=1} = \int d^4 x \sqrt{-g} \left[ \frac{1}{2} \,R - {\cal G}_{i\bar{j}} D_\mu \phi^{i} D^\mu \bar{\phi}^j
- V(\phi_i, \bar{\phi}_i) \right]\,.
\label{eq:action}
\end{equation}
The scalar potential depends on the superpotential $W$, the K\"ahler potential $K$ as well as the K\"ahler metric ${\cal G}_{i\bar{j}} $,
\begin{eqnarray}
V(\phi_i, \bar{\phi}_i)  &=& e^{\mathcal{K}} \left({\cal G}^{i \bar{j}} D_i {W} D_{\bar{j}} \bar{{W}} -3 {W}
\bar{{W}} \right) + 
V_{\text{uplift}} \label{eq:potential}
\\
D_i {W} &=& \partial_i {W} + {W} \partial_i \mathcal{K} \\
{\cal G}_{i\bar{j}} &=& \partial_i \partial_{\bar j} \mathcal{K}
\end{eqnarray}
The derivatives $\partial_i$ and $\partial_{\bar{i}}$ differentiate with respect to the $\phi_i$ and $\bar{\phi}_{\bar{i}}$ dependence, respectively.   
By expanding the complex fields in terms of their real and imaginary part, we can relate the supergravity action above, (\ref{eq:action}), to the action, (\ref{action}), discussed in section~\ref{sfp}.
The term $V_{\text{uplift}}$  will  include the effects of supersymmetry breaking arising from other sectors of the theory.

We will demonstrate our methods in the context of Type IIB string theory compactified to four dimensions on a Calabi-Yau orientifold because the scalar potential in this case is well-understood and realistic four-dimensional models can be constructed~\cite{KKLT,vb-pb-jc-fq,CQS,GKP,DouglasDenef,Allenach}.  
After including the leading perturbative and non-perturbative corrections of string theory, the
K\"ahler potential and superpotential are given by
\begin{eqnarray}
\mathcal{K} & = & - 2 \ln \left(\mathcal{V} + \frac{ \xi \, g_s^{\frac{3}{2}}}{2 e^{\frac{3 \phi}{2}}} \right)
- \ln(-i(\tau - \bar{\tau})) - \ln \left(-i \int_{M} \Omega \wedge \bar{\Omega}\right), \nonumber \\
{W}  & = & \frac{g_s^{\frac{3}{2}} }{\sqrt{4 \pi}} \left(\frac{1}{l_s^2} \int_{M}
G_3 \wedge \Omega + \sum A_i e^{-a_i T_i} \right)
\label{eq:potentials}
\end{eqnarray}
Here $g_s$ is the string coupling, $l_s$ is the string length,  $\Omega$ is the holomorphic three-form on the Calabi-Yau manifold $M$,  $G_3$ is the background field (flux) that is chosen to thread 3-cycles in $M$ and
\be
\xi = -\frac{\zeta(3) \, \chi(M)}{2 (2 \pi)^3}
\label{xidef}
\ee
where $\chi$ is Euler number of $M$.  The axion-dilaton field is $\tau=C_0+ i \, e^{-\phi}$, and the integrals involving $\Omega$ are implicitly functions of the complex structure moduli.  The fields  $T_i = \tau_i + i b_i$ are the complexified K\"ahler moduli where $\tau_i$ is a  4-cycle volume (of the divisor $D_i\in H_4(M,\mbb{Z})$)  and $b_i$ is its axionic partner arising ultimately from the 4-form field.    Here $a_i = 2\pi/N_i$ for some integer $N_i$, for each field, that is determined by the dynamical origin of the exponentials in the superpotential ($N_i = 1$ for brane instanton contributions, $N_i > 1$ for gaugino condensates). Finally, $\mathcal{V}$ is the dimensionless classical volume of the compactification manifold $M$ (in Einstein frame, but measured in units of the string length).   In terms of the K\"ahler class $J=\sum_i t^i D_i$ (by Poincar\'{e} duality $D_i\in H^2(M,\mbb{Z}$)), with the $t^i$ measuring the areas of 2-cycles, $C_i$,
\begin{equation}
{\mathcal V} =  \int_M J^3 = \frac{1}{6} \kappa_{ijk} t^i t^j t^k~,
\label{CYvolume}
\end{equation}
where $\kappa_{ijk}$ are the intersection numbers of the manifold.   ${\mathcal V}$ should be understood as an implicit function of the complexified 4-cycle moduli $T_k$ via the relation
\begin{equation}
\tau_i = \partial_{t_i} {\mathcal V} = \frac{1}{2} \kappa_{ijk} t^j t^k~.
\label{4to2cycles}
\end{equation}

There are additional perturbative corrections to $\mathcal{K}$ in (\ref{eq:potentials}), but we have kept the terms that give the leading contributions to the scalar potential in the large $\mathcal{V}$ limit of interest to us \cite{bergetal}.  In particular, expanding $\mathcal{K}$ to linear order in $\xi$ gives a consistent approximation in inverse powers of $\mc{V}$.   We have also assumed that all of the K\"ahler moduli $T_i$ appear in the superpotential (see \cite{DouglasDenef} for examples) and that we use a basis of 4-cycles such that the exponential terms in ${W}$ take the form $exp(- a_i \, T_i)$.   As these exponentials arise from an instanton expansion, in order to only keep the first term as we have done, the 4-cycle volumes must be sufficiently large to ensure that $a_i T_i \gg 1$. 

Finally, the form of the term $V_{\text{uplift}}$ in (\ref{eq:potential}) depends on the kind of supersymmetry breaking effects that arise from other sectors of the theory.  
We take
\begin{equation}
V_{\text{uplift}} = \frac{\gamma}{\mc{V}^2}\,
\label{uplift}
\end{equation}
which will describe the energy of a space-filling antibrane \cite{KKLT}, fluxes of gauge fields living on D7-branes \cite{BKQ}, or the F-term due to a non-supersymmetric solution for the complex structure/axion-dilaton moduli \cite{Saltman:2004sn}.

It was shown in \cite{KKLT} that a generic choice of background fields $G_3$ causes all the complex structure moduli and the axion-dilaton to acquire string scale masses without breaking supersymmetry.  They are then decoupled from the low-energy theory and their contributions to $\mc{K}$ and ${W}$ are constants for our purposes\footnote{In the case of the F-term breaking due to the complex structure/axion-dilaton moduli \cite{Saltman:2004sn}, the contribution of the complex structure and axion-dilaton moduli to the scalar potential does depend on the volume \eqref{uplift}.}:
\begin{eqnarray}
\mc{K} & = & - 2 \, \ln \left(\mc{V} + \frac{ \xi}{2 } \right)
-\ln \left( \frac{2}{g_s}\right) + \mc{K}_0 , \nonumber \\
{W}  & = & \frac{g_s^{\frac{3}{2}} }{\sqrt{4 \pi} } \left( W_0 + \sum_i A_i e^{-a_i T_i} \right)\,,
\label{eq:potentials2}
\end{eqnarray}
where ${\cal K}_0$ ($W_0$) is the complex structure K\"ahler potential (superpotential), evaluated at the locations where the complex structure moduli have been fixed.
It was shown in \cite{vb-pb-jc-fq} that, when the Euler number, $\chi < 0$,  for generic values of $W_0$ (and hence of the background fluxes $G_3$), the scalar potential for the K\"ahler moduli has a minimum where the volume ${\mc V}$  of the Calabi-Yau manifold $M$ is very large -- the associated energy scale is a few orders of magnitude lower than the GUT scale.     Furthermore, in these Large Volume Scenarios  there is a natural hierarchy -- one of the K\"ahler moduli is much larger than the others and dominates the volume of the manifold.  
%Because the Large Volume Scenarios exist for generic choices of fluxes (unlike the KKLT scenarios \cite{KKLT} which require fine-tuning) they are the statistically favored setting for Type IIB model building.   
For our purposes they are also attractive because the scalar potential admits an expansion in inverse powers of the large volume $\mc{V}$.  This will allow us to carry out analytical calculations of inflation arising from K\"ahler moduli rolling towards the large volume minimum of the potential.

Several previous works have considered inflation in the large volume setting, e.g., \cite{0509012,roulette,ourprevious,jimmy}. 
Here we include all K\"ahler moduli, and not just the light modes. Although we find that the heavy modes, corresponding to K\"ahler moduli that are stabilized before inflation takes, do not affect the dynamics during inflation
in the models that we have studied, these modes do change after inflation has ended.

Slow roll inflation can occur in  a region of the field space where the potential is positive and very flat.  We will look for this in the Large Volume Scenarios described above, where, at the minimum of the scalar potential, there is a hierarchy amongst the K\"ahler moduli
\begin{equation}
\tau_1 \gg \tau_2, \tau_{3} , \tau_4 \cdots
\label{modulihierarchy}
\end{equation}
which we will use to simplify the effective potential. 

For transparency of the equations, we will assume  that the intersection numbers $k_{ijk}$ are such that in the basis of 4-cycles, $\tau_i$, the volume takes the diagonal form~\cite{CQS} 
\begin{equation}
\mathcal{V} = \alpha ({\tau_1}^{\frac{3}{2}}-\sum_{i=2} \lambda_i {\tau_i}^{\frac{3}{2}}) = - \alpha\sum_{i=1} \lambda_i {\tau_i}^{\frac{3}{2}}
\end{equation}
where  $\lambda_1 = -1$, and $\lambda_i, i \geq 2$ are usually positive. 
%We let the first modulus dominate the volume 
%\[\tau_1 \gg 1, \tau_1 \gg \tau_i, \quad i \geq 2\]
%and thus ignore the exponentials that depend on $\tau_1$

With the volume taking the above form we can explicitly compute the metric on the moduli space, 
${\cal G}_{i \bar{j}} = \partial_i \partial_{\bar{j}} \mathcal{K}$, which is needed both for the 
metric, $h_{ij}$,  and for the scalar potential, $V$, appearing in the four dimensional action~(\ref{action}).
By expanding
in inverse powers of ${\cal V}$, keeping terms to $O({\cal V}^{-2})$, we obtain
\begin{equation}\label{metric}
{\cal G}_{i \bar{j}}= \frac{3\alpha \lambda_i}{8 (\mathcal{V} + \frac{\xi}{2}) {\tau_i}^{\frac{1}{2}}} \delta^{ij} + \frac{ 9 \alpha^2 \lambda_i \lambda_j \sqrt{ \tau_i \tau_j} }{8 (\mathcal{V} + \frac{\xi}{2})^2}\,.
\end{equation}
%Starting with the metric on the moduli space from the supergravity action~(\ref{eq:action}), ${\cal G}_{i \bar{j}} = \partial_i \partial_{\bar{j}} \mathcal{K}$, we can write down the explicit metric, $h_{ij}$, as it appears in the four dimensional action~(\ref{action}), in an expansion in inverse powers of ${\cal V}$ keeping terms to $O({\cal V}^{-2})$,
%%\begin{equation}
%%G_{a \bar{b}} = \partial_a \partial_{\bar{b}} \mathcal{K}
%%\end{equation}
%%written explicitly,
%\begin{equation}
%{\cal G}_{i \bar{j}}= \frac{3\alpha \lambda_i}{8 \mathcal{V} {\tau_i}^{\frac{1}{2}}} \delta^{ij} + \frac{ 9 \alpha^2 \lambda_i \lambda_j \sqrt{ \tau_i \tau_j} }{8 \mathcal{V}^2}
%\end{equation}
With the axions minimized in the potential, %i.e., $cos (a_i\theta_i )= -1$, 
the effective potential then becomes~\cite{0509012}
\begin{equation}\label{effpotential}
V= \sum^4_{i=2}\frac{8(a_i A_i)^2\sqrt{\tau_i}}{3\mathcal{V}\lambda_i \alpha} e^{-2 a_i \tau_i} -  \sum^4_{i=2}\frac{4 a_i A_i W_0 \tau_i}{{\mathcal{V}}^2} e^{- a_i \tau_i} + \frac{3\xi {W_0}^2}{4 {\mathcal{V}}^3} + \frac{ \gamma}{{\mathcal{V}}^2}\,,
\end{equation}
where we have assumed that ${\cal K}_0$ can be chosen such that the overall scale of the potential is simplified, i.e., overall factors of $g_s$ and $2\pi$ are not present.
Here  we have expanded $V$ to $O({\cal V}^{-3})$ to include the leading $\alpha'$-corrections, $\frac{3\xi {W_0}^2}{4 {\mathcal{V}}^3}$, as well as the uplift term, $\frac{ \gamma}{{\mathcal{V}}^2}$. The parameters in the potential can be chosen and tuned under certain constraints~\cite{roulette}\cite{ourprevious}\cite{08092982}. 
%Note that the kinetic term
%\[{\cal G}_{i\bar{j}}\partial_{\mu} T^i \partial^{\mu} \bar{T}^{\bar{j}} = {\cal G}_{i{j}} \partial_{\mu} \tau^i \partial^{\mu} \tau^j\]
%which identifies  ${\cal G}_{a\bar{b}}$  as $h_{ab}$ in~\Eq{action}.

To determine the local minimum (vacuum) of the potential we need to solve the equations 
\begin{equation}\label{localminimum}
\frac{\partial V}{\partial \tau_i} = 0
\end{equation}
While it is difficult to get the analytical results\footnote{Altough one can make approximations to solve the minimum equations analytically as in \cite{0509012} and \cite{08043653}, it is desirable to solve them numerically. As we can show by numerical analysis, the analytical solutions after approximation will likely spoil the results.}, these equations can always be solved numerically.

It is more convenient to work in the canonical frame, rather than the form taken by the supergravity metric in Eq.~\Eq{metric}, since we have already solved the perturbation equations in the canonical frame\footnote{Note that comparing the kinetic energy terms in the actions~\Eq{action} and~\Eq{eq:action}, respectively, we find that $h_{ij} = 2{\cal G}_{i\bar j}$, with ${\cal G}_{i\bar j}$ given in~\Eq{metric}.}.
Although it is difficult to find the exact transformations which can diagonalize the metric, we do find a canonical frame which is a good approximation as long as $\tau_1 \gg \tau_i$, of which the field space transformations are
\begin{align}
\phi^1 &= \sqrt{\frac{3\lambda_1(1 + 3\lambda_1)}{4}}\, log(\tau_1) \label{fieldtran1}\\
\phi^i  &= \sqrt{ \frac{4 \lambda_i}{3 {\tau_1}^{\frac{3}{2}} }}{\tau_i}^{\frac{3}{4}}, \text{ } i\geq 2\label{fieldtran2}
\end{align}
%\begin{align}
%\phi^1 &= \sqrt{\frac{3\lambda_1(1 + 3\lambda_1)}{8}} log(\tau_1) \label{fieldtran1}\\
%\phi^i  &= \sqrt{ \frac{2 \lambda_i}{3 {\tau_1}^{\frac{3}{2}} }}{\tau_i}^{\frac{3}{4}}, \text{ } i\geq 2\label{fieldtran2}
%\end{align}
During, as well as after inflation, the metric, in terms of the above redefined fields, $\phi^i$, remains canonically normalized, to leading order in inverse powers of the volume. Considering that the original metric, ${\cal G}_{i\bar j}$ is a K\"ahler metric which neither is field independent nor diagonal, this result is somewhat surprising.

\section{Model Study}
\label{ms}
In general, a multi-field inflationary model should contain both the heavy fields and the light fields\footnote{As has been shown in \cite{0906.3711}, the fields that are heavy (light) during inflation may become light (heavy) after inflation ends. So the heaviness (or the lightness) of a field is determined not only  by the corresponding parameters, but to a large extent also by its position/value in the field space.}.  To obtain inflation we choose the initial conditions such that the light fields are displaced away from the local minimum and the heavy fields are at the corresponding local minimum once the initial values of the light fields are chosen. We expect that the heavy fields will be frozen as the light fields approach the minimum.  As we will see later in the numerical analysis, the light fields carry all the kinetic energy and are responsible for the creation of inflation. The number of e-foldings or the duration of inflation is determined by how far away the light fields are displaced from the minimum. The heavy fields will only begin to move and oscillate together with the light fields around the local minimum shortly after the end of inflation.

In what follows, we will discuss two example models based on the discussion in the previous section. In both cases there are two heavy fields/moduli. The former has a single light field (inflaton) and the latter has two. By assigning appropriate values to the parameters in the effective potential, we solve the background equations of motion numerically. Next, we perform the field transformation~\Eq{fieldtran1}, \Eq{fieldtran2} to get the the kinetic energy in its canonical form. Then we use the perturbation solutions(light) to compute the curvature and isocurvature perturbations. Finally, we calculate the spectra and tilts at Hubble exit. Our models can be easily reduced or generalized. 

\subsection{The Three-Field Model}
Let us construct an inflationary model with two heavy moduli, $\tau_1$ and $\tau_2$, and a light modulus, $\tau_3$. 
This is essentially the Conlon-Quevedo model~\cite{0509012}. However, we do not assume that the initial values of the heavy moduli are the same as the final values after inflation.  

The parameters in the effective potential~\Eq{effpotential} are set to be 
\[  \alpha=\frac{1}{9\sqrt{2}}, a_2=\frac{2\pi}{300}, a_3=\frac{2\pi}{100},  A_2=0.2, A_3=0.002 \]
\[\lambda_1 = -1,  \lambda_2=0.1 , \lambda_3=0.010, W=500, \xi=40, \gamma=9.75 \times 10^{-6}  \]
With these parameters, the local minimum is at
\[\tau_{1min}=62100.7, \tau_{2min}=234.1, \tau_{3min}= 69.0202 \]

The initial conditions  imposed are~\footnote{The attractor behavior of the evolution equations will ensure the same terminal velocity shortly after inflation begins regardless of how we choose the initial velocities~\cite{arXiv:astro-ph/0703486}. }
\[\tau_{1}(0)=76212.1, \tau_{2}(0)=246.99, \tau_{3}(0)= 472, \dot{\tau_{1}}(0) =\dot{\tau_{2}}(0) = 0,, \dot{\tau_{3}}(0) =-7.13\times10^{-19}\]
Obviously, all $\tau_{i}(0)$'s are quite different from $\tau_{imin}$'s.
\begin{figure}[ht]
\begin{center}
\begin{tabular}{ll}
\scalebox{0.60}{\includegraphics*{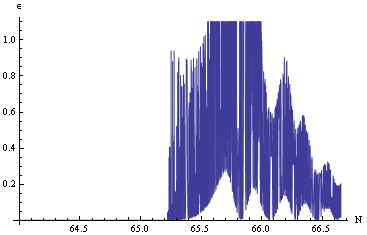}}&\, 
\scalebox{0.65}{\includegraphics*{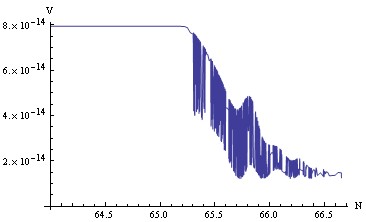}} \\
\hskip 3.1truecm{\small (a)}            &
\hskip 3.1truecm{\small (b)}\\
\end{tabular}
\end{center}
\caption{\small{Inflation in the last few e-foldings.
a) The slow-roll parameter $\epsilon$.
b) The potential $V$.}}
\label{ep31-V3}
\end{figure}

\begin{figure}[ht]
\begin{center}
\begin{tabular}{ll}
\scalebox{0.62}{\includegraphics*{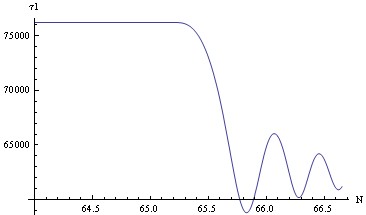}}&\, 
\scalebox{0.62}{\includegraphics*{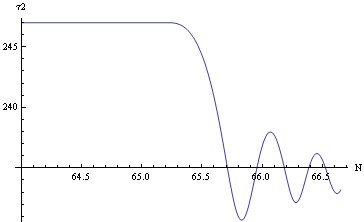}} \\
\hskip 3.1truecm{\small (a)}            &
\hskip 3.1truecm{\small (b)}\\
\end{tabular}
\end{center}
\caption{\small{Evolution of the heavy fields in the last few e-foldings.
a) $\tau_1$.
b) $\tau_2$.}}
\label{tau13-tau23}
\end{figure}

\begin{figure}
\begin{center}
\scalebox{0.7}{\includegraphics*{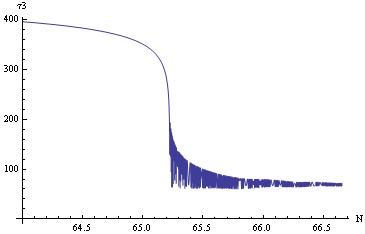}} 
\end{center}
\caption{Evolution of the light field $\tau_3$ in the last few e-foldings.}
\label{tau33}
\end{figure}

Apparently, the heavy moduli are frozen until the end of inflation, and $\tau_{\text{final}}\to \tau_{min}$. All the moduli began to oscillate after inflation ends. The light moduli oscillate much faster than the heavy moduli.  The generated inflation lasts about $N_{tot}=65.2$ e-foldings. We calculate  the spectral index for the curvature perturbation at $N_*=60$ e-foldings (counted backward from the end of infaltion) before the end of inflation
\[n_{\mathcal{R}}=n_{\mathcal{R}_*} = 0.9563\, (\text{the running }\tilde{\alpha} = 0), \epsilon= 2.69 \times 10^{-13}\]
Note that the spectral index is calculated at the Hubble exit (denoted by $_*$). In the calculation (by~\Eq{apsd}), we consider perturbations of both the heavy moduli and the light modulus. The same will be done in the next section.

\subsection{The Four-Field Models}
\paragraph{Symmetric Case}
First, we want to have two identical/symmetric light fields for which both the parameters and the initial conditions are the same. We expect to get a straight line background trajectory.
The parameters in the effective potential~\Eq{effpotential} are set to be 
%The parameters 
\[  \alpha=\frac{1}{9\sqrt{2}}, a_2=\frac{2\pi}{300}, a_3=\frac{2\pi}{100},  A_2=0.2, A_3=0.001, A_4=0.001 \]
\[\lambda_1 = -1,  \lambda_2=0.1 , \lambda_3=0.005, \lambda_4=0.005, W=500, \xi=40, \gamma=9.75 \times 10^{-6}  \]
With these parameters, the local minimum is at
%The local minimum 
\[\tau_{1min}=62100.7, \tau_{2min}=234.1, \tau_{3min}= 69.0202, \tau_{4min}= 69.0202 \]

The initial conditions imposed are
\[\tau_{1}(0)=76212.1, \tau_{2}(0)=246.99, \tau_{3}(0)= 472, \tau_{3}(0)= 472, \]
\[\dot{\tau_{1}}(0) =\dot{\tau_{2}}(0) = 0, \dot{\tau_{3}}(0) =\dot{\tau_{4}}(0) =-1.71\times10^{-19}\]

\begin{figure}[ht]
\begin{center}
\begin{tabular}{ll}
\scalebox{0.60}{\includegraphics*{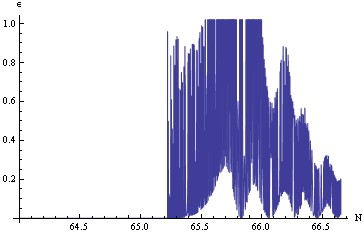}}&\, 
\scalebox{0.65}{\includegraphics*{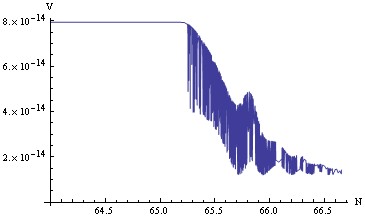}} \\
\hskip 3.1truecm{\small (a)}            &
\hskip 3.1truecm{\small (b)}\\
\end{tabular}
\end{center}
\caption{\small{
a) The slow-roll parameter $\epsilon$.
b) The potential $V$.}}
\label{ep41-V4}
\end{figure}

\begin{figure}[ht]
\begin{center}
\begin{tabular}{ll}
\scalebox{0.62}{\includegraphics*{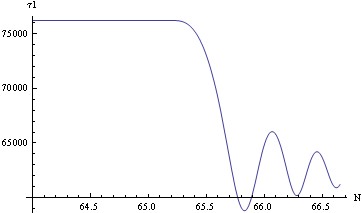}}&\, 
\scalebox{0.62}{\includegraphics*{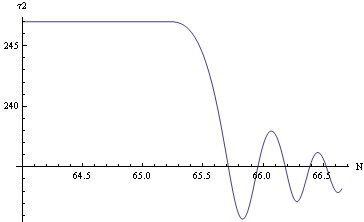}} \\
\hskip 3.1truecm{\small (a)}            &
\hskip 3.1truecm{\small (b)}\\
\end{tabular}
\end{center}
\caption{\small{Evolution of the heavy fields in the last few e-foldings.
a) $\tau_1$.
b) $\tau_2$.}}
%\label{tau13-tau23}
\end{figure}
\begin{figure}[ht]
\begin{center}
\begin{tabular}{ll}
\scalebox{0.62}{\includegraphics*{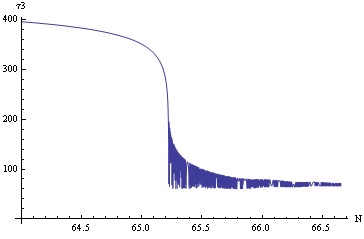}}&\, 
\scalebox{0.62}{\includegraphics*{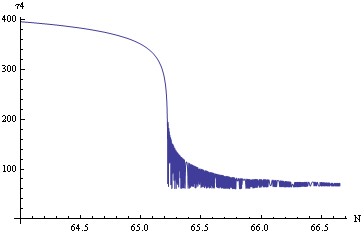}} \\
\hskip 3.1truecm{\small (a)}            &
\hskip 3.1truecm{\small (b)}\\
\end{tabular}
\end{center}
\caption{\small{Evolution of the light fields in the last few e-foldings.
a) $\tau_3$.
b) $\tau_4$.}}
%\label{tau13-tau23}
\end{figure}

Again, the heavy moduli remain frozen until after inflation ends. The generated inflation lasts about $N_{tot}\approx 65.2$. The spectral indices, computed as in the example above at $N_*=60$ e-foldings before the end of inflation, are given by
\bea
n_{\mathcal{R}_*}&{=}&0.9621\, (\tilde{\alpha}_{\mathcal{R}} {=} -1.2\times 10^{-15}),\\
n_{\mathcal{S}_*}&{=}&0.9621\, (\tilde{\alpha}_{\mathcal{S}} {=} -1.2\times 10^{-15}),\\
 cos\Delta_*&{=}&-1.5\times 10^{-5}, \epsilon{=} 2.69 \times 10^{-13}\,.
\eea
The spectral indices at Hubble crossing satisfies $n_{\mathcal{R}_*} =n_{\mathcal{S}_*}$ because we have a symmetric model.  The correlation angle $cos\Delta_*$ is close to zero, and it is consistent with \cite{0605679} in which  
\[cos\Delta_* \simeq - 2 C {\eta_{\sigma s}}_* = - 2 C cos\theta^I \beta^J V_{IJ}|_* = -1.98\times 10^{-5} \]
where $C \simeq 0.7296$.

\paragraph{Nonsymmetric Case}
This time, we have two light fields of which  the parameters are identical but the initial conditions are different. 
The parameters in the effective potential~\Eq{effpotential} are set to be 
\[  \alpha=\frac{1}{9\sqrt{2}}, a_2=\frac{2\pi}{300}, a_3=\frac{2\pi}{100},  A_2=0.2, A_3=0.001, A_4=0.001 \]
\[\lambda_1 = -1,  \lambda_2=0.1 , \lambda_3=0.005, \lambda_4=0.005, W=500, \xi=40, \gamma=9.75 \times 10^{-6}  \]
\[\tau_{1min}=62100.7, \tau_{2min}=234.1, \tau_{3min}= 69.0202 \]

The initial conditions are
\[\tau_{1}(0)=76212.1, \tau_{2}(0)=246.99, \tau_{3}(0)= 472, \tau_{3}(0)= 492,\]
\[\dot{\tau_{1}}(0) =\dot{\tau_{2}}(0) = 0, \dot{\tau_{3}}(0) =-1.72\times10^{-19}, \dot{\tau_{4}}(0) =-1.5\times10^{-19}\]

\begin{figure}[ht]
\begin{center}
\begin{tabular}{ll}
\scalebox{0.60}{\includegraphics*{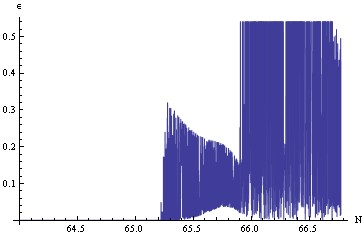}}&\, 
\scalebox{0.65}{\includegraphics*{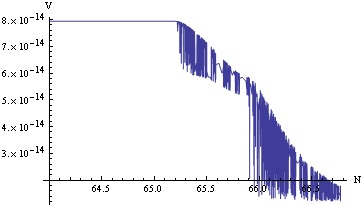}} \\
\hskip 3.1truecm{\small (a)}            &
\hskip 3.1truecm{\small (b)}\\
\end{tabular}
\end{center}
\caption{\small{
a) The slow-roll parameter $\epsilon$.
b) The potential $V$.}}
\label{ep41p-V4p}
\end{figure}

\begin{figure}[ht]
\begin{center}
\begin{tabular}{ll}
\scalebox{0.62}{\includegraphics*{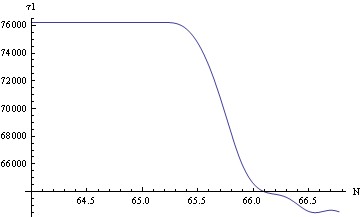}}&\, 
\scalebox{0.62}{\includegraphics*{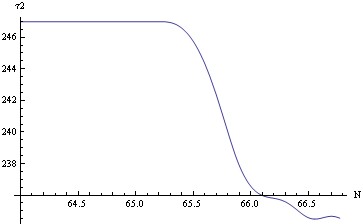}} \\
\hskip 3.1truecm{\small (a)}            &
\hskip 3.1truecm{\small (b)}\\
\end{tabular}
\end{center}
\caption{\small{Evolution of the heavy fields in the last few e-foldings.
a) $\tau_1$.
b) $\tau_2$.}}
\label{tau13p-tau23p}
\end{figure}
\begin{figure}[ht]
\begin{center}
\begin{tabular}{ll}
\scalebox{0.62}{\includegraphics*{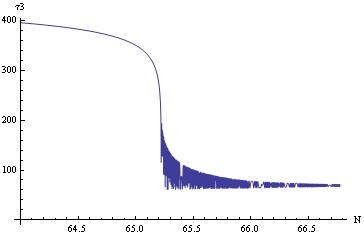}}&\, 
\scalebox{0.62}{\includegraphics*{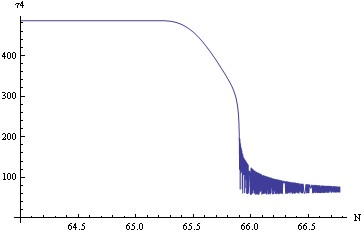}} \\
\hskip 3.1truecm{\small (a)}            &
\hskip 3.1truecm{\small (b)}\\
\end{tabular}
\end{center}
\caption{\small{Evolution of the light fields in the last few e-foldings.
a) $\tau_3$.
b) $\tau_4$.}}
%\label{tau13p-tau23p}
\end{figure}

The generated inflation lasts about $N_{tot}\approx 66$. In this example, strictly speaking, inflation does not end initially when the inflaton($\tau_3$) begins to oscillate. It ends when other fields also begin to oscillate. 

 The spectral indices  at $N_*=60$ e-foldings before the end of inflation are found to be
 \bea
%\[
n_{\mathcal{R}_*}&{=}&0.9639\,(\tilde{\alpha}_{\mathcal{R}} {=} 4.69\times 10^{-5}),\,\\
 n_{\mathcal{S}_*}&{=}&0.9879\,(\tilde{\alpha}_{\mathcal{S}} {=} 4.85\times 10^{-5}),\, \\
 cos\Delta_* &{=}& -0.00501, \epsilon{=} 1.44 \times 10^{-13}\,.
 \eea

 Again, the correlation angle $cos\Delta_*$ is very small, consistent with \cite{0605679} where  
\[cos\Delta_* \simeq - 2 C {\eta_{\sigma s}}_* = - 2 C cos\theta^I \beta^J V_{IJ}|_* =0.00507 \]

\section{Conclusions}
\label{conc}
In the typical inflation scenery, some or all of the light fields act as inflatons which are initially displaced from their local minimum. Along the path of the light fields the potential is very flat, and the light fields will slowly roll to their local potential minimum during inflation.  Their perturbations are almost invariant and only decay slowly as the universe expands. 

The heavy fields, on the other hand, should be frozen during inflation, until shortly after the end of inflation when they start to move from the initial local minimum to the final local minimum. Their perturbations (as shown by the solution~\Eq{solution_perturbation_approx_l_im_ord_s_arg}) decay rapidly, approximately as $a^{-3}$, due to the expansion of the universe. That is why we ignore the heavy solutions (those with $-\lambda_I >1$) when we calculate the perturbations analytically.

The resulting power spectrum of the curvature perturbation can be calculated using~\Eq{solpem0}, which is almost scale invariant due to the small $\lambda_I$.  If there is only one light field, the contribution from the perturbations of the light fields is purely adiabatic. The perturbations are constants during inflation and reheating~\cite{tasi}. If there is more than one light field, they will also generate perturbations orthogonal to the background trajectory. In this case,  the presence of the non-adiabatic/entropy perturbations can change the curvature perturbations, disqualifying it as a constant of time as discussed in section~3.

The ratio of the curvature perturbation and isocurvature perturbation is
\begin{equation}\label{correlation-iso-ad}
\alpha = \frac{P_S}{P_R + P_S},
\end{equation}
Observational constraints from WMAP temperature anisotropy favor a small $\alpha$ with an upper limit 0.070(95\% CL)~\cite{08123500}. The single (light) field inflation always satisfies this limit since there is no isocurvature perturbation. The multi-light-filed inflation can also satisfy this limit if the transfer functions can efficiently suppress the amplitude of the isocurvature perturbation after Hubble  crossing. As discussed in section~3.2, the transfer functions depend on the details of the physics during as well as after the primordial era, and we hope to address this issue in the near future. 

We discussed an interesting inflation model derived from string compactification in the last two sections. Its potential has a nice shape which is ideal for generating inflation. The result is very attractive. By adjusting the parameters and assigning appropriate initial conditions, we get the required number of e-foldings before the end of inflation.  The numerical analysis shows that there is an oscillating period after the end of inflation which should be identified with the ``preheating'' era. It shows that the light fields oscillate much faster than the heavy fields. The preheating lasts a few e-foldings before the moduli roll to the minimum (vacuum) in the potential.  In the single light field model, the spectral index calculated at $N_*=60$ e-foldings before the end of inflation is about 0.956,  which is consistent with the five-year WMAP data, $n_s = 0.960 \pm 0.013$~\cite{arXiv:0803.0547}. In the multi-light fields model, we get ${n_s}_* \sim 0.962-0.964$. However, in this case, it will require a better understanding of the transfer function, which is one of the main tasks left for future investigation, before we can compare the results with observations. 

\acknowledgments

We thank V.~Balasubramanian, R.~Jimenez, J.~Simon and L.~Verde for collaborations on earlier
papers that lead us to this study. P.B. is also grateful to  J.J.~Blanco-Pillado for interesting discussions
related to the paper~\cite{0906.3711}.
This work has been carried out with support from the NSF CAREER grant PHY-0645686.
P.B. acknowledges additional support from
the University of New Hampshire through its Faculty Scholars Award
Program.
%\vfil\eject
\section*{Appendix}  % use *-form to suppress numbering
\appendix

\section{The Hankel Function of Large Complex Order}
\renewcommand{\theequation}{A-\arabic{equation}}
% redefine the command that creates the equation no.
\setcounter{equation}{0}  % reset counter 
First, we use the Frobenius expansion of the Bessel function around the origin

\begin{align} 
J_\mu (z )& = \left(\frac{z}{2}\right)^\mu\left[\frac{1}{\Gamma (\mu +1)}-\frac{1}{\Gamma(\mu+2)}{\left(\frac{z}{2}\right)}^2+\frac{1}{2!\Gamma (\mu +3)} \left(\frac{z}{2}\right)^4-\cdots\right] \nonumber \\
&= \left(\frac{z}{2}\right)^\mu \sum_{k=0}^\infty (-1)^k\frac{(z/2)^{2k}}{\Gamma(1+\mu+k)k!}, \text{ } z \ll 1
\end{align}
To the lowest order,
\begin{equation} 
\left\{ J_\mu \simeq \frac{(z/2)^\mu}{\Gamma(1+\mu)}, J_{-\mu} \simeq \frac{(z/2)^{-\mu}}{\Gamma(1-\mu)} \right\} , \text{ } z \ll 1
\end{equation}

So 
\begin{equation}\label{hankel_zero_exp}
H^{(1)}_\mu (z) = \frac{e^{-i\mu\pi} J_\mu (z ) - J_{-\mu}(z)} {-i\sin\pi\mu} 
\simeq \frac{e^{-i\mu\pi} \frac{(z/2)^\mu}{\Gamma(1+\mu)} - \frac{(z/2)^{-\mu}}{\Gamma(1-\mu)}} {-i\sin\pi\mu}
\end{equation}

Stirling's formula for large $\nu$ approximation of the Gamma function gives 
\begin{equation}\label{Gamma_Striling}
\Gamma(\nu) = \nu^\nu e^{-\nu} \sqrt{\frac{2\pi}{\nu}}[1+ \frac{1}{12\nu} + \frac{1}{288\nu^2} + O(\nu^{-3})], \text{ }|\nu| > 1, \text{ } |arg(\nu)| < \pi-\epsilon
\end{equation}
where $\epsilon$ is any small positive number. 

Let $\nu =  1 \pm i\rho$ where $\rho >1$,  to lowest order, 
\begin{align}
\Gamma(1 \pm i\rho) &\approx (1 \pm i\rho )^{1\pm i\rho} e^{-1 \mp i\rho} \sqrt{\frac{2\pi}{1\pm i\rho}} \nonumber \\ 
& = (1 \pm i\rho )(1 \pm i\rho )^{\pm i\rho}e^{-1}e^{\mp i\rho}\sqrt{2\pi} \frac{1}{(1+\rho^2)^{\frac{1}{4}}} e^{\mp i \frac{\alpha}{2}} \nonumber \\
& =\sqrt{2\pi}(1+\rho^2)^{\frac{1}{4}} e^{-1 -\alpha \rho} e^{\pm i(\frac{\alpha}{2} -\rho + \rho log\sqrt{1 + \rho^2})}
\end{align}
where $1 \pm i \rho = \sqrt{1+\rho^2}e^{\pm i \alpha}$.

So for $\rho > 1$, the Hankel function is 
\begin{equation}\label{Hankel_approx_l_im_ord_s_arg}
H^{(1)}_{\mu} (z) \approx \begin{cases}
\sqrt{\frac{2}{\pi}} (1+\rho^2)^{-\frac{1}{4}} e^{1 + \alpha\rho } \omega, \quad &\mu= i\rho \\
\sqrt{\frac{2}{\pi}} (1+\rho^2)^{-\frac{1}{4}} e^{1 + (\alpha-\pi)\rho }\omega,  \quad &\mu= -i\rho
\end{cases}
\end{equation}
where 
\begin{equation}
\omega =e^{-\rho \pi} (\frac{z}{2})^{-i\rho} e^{i(\frac{\alpha}{2} -\rho + \rho log\sqrt{1 + \rho^2})} - (\frac{z}{2})^{i\rho}e^{-i(\frac{\alpha}{2} -\rho + \rho log\sqrt{1 + \rho^2})}
\end{equation}


\begin{thebibliography}{77}

\bibitem{guth1981}
  A.~H.~Guth,
  ``The Inflationary Universe: A Possible Solution To The Horizon And Flatness
  Problems,''
  Phys.\ Rev.\  D {\bf 23}, 347 (1981).
  %%CITATION = PHRVA,D23,347;%%
\bibitem{Linde:1981mu}
A.~D.~Linde,
``A New Inflationary Universe Scenario: A Possible Solution Of The Horizon,
Flatness, Homogeneity, Isotropy And Primordial Monopole Problems,''
Phys.\ Lett.\  B {\bf 108}, 389 (1982).
%%CITATION = PHLTA,B108,389;%%

\bibitem{Albrecht:1982wi}
A.~J.~Albrecht and P.~J.~Steinhardt,
``Cosmology For Grand Unified Theories With Radiatively Induced Symmetry
Breaking,''
Phys.\ Rev.\ Lett.\  {\bf 48}, 1220 (1982).
%%CITATION = PRLTA,48,1220;%%

\bibitem{lythriotto1999}
  D.~H.~Lyth and A.~Riotto,
  ``Particle physics models of inflation and the cosmological density
  perturbation,''
  Phys.\ Rept.\  {\bf 314}, 1 (1999)
  [arXiv:hep-ph/9807278].
  %%CITATION = PRPLC,314,1;%%

\bibitem{tasi}
  D.~Baumann,
  ``TASI Lectures on Inflation,''
  arXiv:0907.5424 [hep-th].
  %%CITATION = ARXIV:0907.5424;%%

\bbibitem{KKLT}
S.~Kachru, R.~Kallosh, A.~Linde and S.~P.~Trivedi,
``De Sitter vacua in string theory,''
Phys.\ Rev.\  D {\bf 68}, 046005 (2003)
[arXiv:hep-th/0301240].
%%CITATION = PHRVA,D68,046005;%%

\bibitem{vb-pb-jc-fq}
  V.~Balasubramanian, P.~Berglund, J.~P.~Conlon and F.~Quevedo,
  ``Systematics of Moduli Stabilisation in Calabi-Yau Flux Compactifications,''
  JHEP {\bf 0503}, 007 (2005)
  [arXiv:hep-th/0502058].
  %%CITATION = JHEPA,0503,007;%%

\bibitem{vb-pb}
  V.~Balasubramanian and P.~Berglund,
  ``Stringy corrections to Kahler potentials, SUSY breaking, and the
  cosmological constant problem,''
  JHEP {\bf 0411}, 085 (2004)
  [arXiv:hep-th/0408054].
  %%CITATION = JHEPA,0411,085;%%

\bibitem{0509012}
  J.~P.~Conlon and F.~Quevedo,
  ``Kaehler moduli inflation,''
  JHEP {\bf 0601}, 146 (2006)
  [arXiv:hep-th/0509012].
  %%CITATION = JHEPA,0601,146;%%

\bibitem{08043653}
  H.~X.~Yang and H.~L.~Ma,
  ``Two-field K\'ahler moduli inflation on large volume moduli stabilization,''
  JCAP {\bf 0808}, 024 (2008)
  [arXiv:0804.3653 [hep-th]].
  %%CITATION = JCAPA,0808,024;%%

\bibitem{0906.3711}
  J.~J.~Blanco-Pillado, D.~Buck, E.~J.~Copeland, M.~Gomez-Reino and N.~J.~Nunes,
  %``Kahler Moduli Inflation Revisited,''
  arXiv:0906.3711 [hep-th].
  %%CITATION = ARXIV:0906.3711;%%

%\cite{Burgess:2002ub}
\bibitem{Burgess:2002ub}
 C.~P.~Burgess, J.~M.~Cline, F.~Lemieux and R.~Holman,
 ``Are inflationary predictions sensitive to very high energy physics?,''
 JHEP {\bf 0302}, 048 (2003)
 [arXiv:hep-th/0210233].
 %%CITATION = JHEPA,0302,048;%%


\bibitem{0605679}
  C.~T.~Byrnes and D.~Wands,
  ``Curvature and isocurvature perturbations from two-field inflation in a
  slow-roll expansion,''
  Phys.\ Rev.\  D {\bf 74}, 043529 (2006)
  [arXiv:astro-ph/0605679].
  %%CITATION = PHRVA,D74,043529;%%

\bibitem{mukhanov}
  V.~F.~Mukhanov, H.~A.~Feldman and R.~H.~Brandenberger,
  ``Theory of cosmological perturbations. Part 1. Classical perturbations. Part
  2. Quantum theory of perturbations. Part 3. Extensions,''
  Phys.\ Rept.\  {\bf 215}, 203 (1992).
  %%CITATION = PRPLC,215,203;%%

\bibitem{nast}
  T.~T.~Nakamura and E.~D.~Stewart,
  ``The spectrum of cosmological perturbations produced by a multi-component
  inflaton to second order in the slow-roll approximation,''
  Phys.\ Lett.\  B {\bf 381}, 413 (1996)
  [arXiv:astro-ph/9604103].
  %%CITATION = PHLTA,B381,413;%%

\bibitem{astro-ph/0106020}
  D.~J.~Schwarz, C.~A.~Terrero-Escalante and A.~A.~Garcia,
  ``Higher order corrections to primordial spectra from cosmological
  inflation,''
  Phys.\ Lett.\  B {\bf 517}, 243 (2001)
  [arXiv:astro-ph/0106020].
  %%CITATION = PHLTA,B517,243;%%

\bibitem{0908.0585}
  S.~Hirai and T.~Takami,
  ``Scale dependence of the power spectrum of the curvature perturbation
  determined using a numerical method in slow-roll inflation,''
  arXiv:0908.0585 [astro-ph.CO].


\bibitem{mukhanov-book} V.~F.~Mukhanov, \emph{Physical Foundations of Cosmology}, Cambridge University Press, 2005.

\bibitem{07040212}
  Z.~Lalak, D.~Langlois, S.~Pokorski and K.~Turzynski,
  ``Curvature and isocurvature perturbations in two-field inflation,''
  JCAP {\bf 0707}, 014 (2007)
  [arXiv:0704.0212 [hep-th]].

\bibitem{arXiv:astro-ph/9607038}
  E.~F.~Bunn, A.~R.~Liddle and M.~J.~.~White,
  ``Four-year COBE normalization of inflationary cosmologies,''
  Phys.\ Rev.\  D {\bf 54}, 5917 (1996)
  [arXiv:astro-ph/9607038].
  %%CITATION = PHRVA,D54,5917;%%

\bibitem{arXiv:0803.0547}
  E.~Komatsu {\it et al.}  [WMAP Collaboration],
  ``Five-Year Wilkinson Microwave Anisotropy Probe (WMAP)%\altaffilmark 1 )
  Observations:Cosmological Interpretation,''
  Astrophys.\ J.\ Suppl.\  {\bf 180}, 330 (2009)
  [arXiv:0803.0547 [astro-ph]].
  %%CITATION = APJSA,180,330;%%


\bibitem{08121079}
  S.~Hirai and T.~Takami,
  %``Time dependence of cosmological and inflationary parameters in slow-roll
  %inflation,''
  arXiv:0812.1079 [astro-ph].
  %%CITATION = ARXIV:0812.1079;%%

\bibitem{peiris}
  H.~V.~Peiris {\it et al.}  [WMAP Collaboration],
  ``First year Wilkinson Microwave Anisotropy Probe (WMAP) observations:
  Implications for inflation,''
  Astrophys.\ J.\ Suppl.\  {\bf 148}, 213 (2003)
  [arXiv:astro-ph/0302225].
  %%CITATION = APJSA,148,213;%%

\bibitem{0003278}
  D.~Wands, K.~A.~Malik, D.~H.~Lyth and A.~R.~Liddle,
  ``A new approach to the evolution of cosmological perturbations on large
  scales,''
  Phys.\ Rev.\  D {\bf 62}, 043527 (2000)
  [arXiv:astro-ph/0003278].
  %%CITATION = PHRVA,D62,043527;%%

\bibitem{0205253}
  D.~Wands, N.~Bartolo, S.~Matarrese and A.~Riotto,
  ``An observational test of two-field inflation,''
  Phys.\ Rev.\  D {\bf 66}, 043520 (2002)
  [arXiv:astro-ph/0205253].
  %%CITATION = PHRVA,D66,043520;%%

\bibitem{0107502}
  N.~Bartolo, S.~Matarrese and A.~Riotto,
  ``Adiabatic and isocurvature perturbations from inflation: Power spectra  and
  consistency relations,''
  Phys.\ Rev.\  D {\bf 64}, 123504 (2001)
  [arXiv:astro-ph/0107502].
  %%CITATION = PHRVA,D64,123504;%%

\bbibitem{CQS}
J.~P.~Conlon, F.~Quevedo and K.~Suruliz,
``Large-volume flux compactifications: Moduli spectrum and D3/D7 soft
supersymmetry breaking,''
JHEP {\bf 0508}, 007 (2005)
[arXiv:hep-th/0505076].
%%CITATION = JHEPA,0508,007;%%

\bibitem{9704452}
  L.~Kofman, A.~D.~Linde and A.~A.~Starobinsky,
  ``Towards the theory of reheating after inflation,''
  Phys.\ Rev.\  D {\bf 56}, 3258 (1997)
  [arXiv:hep-ph/9704452].
  %%CITATION = PHRVA,D56,3258;%%


\bibitem{08103913}
  C.~T.~Byrnes,
  ``Constraints on generating the primordial curvature perturbation and
  non-Gaussianity from instant preheating,''
  JCAP {\bf 0901}, 011 (2009)
  [arXiv:0810.3913 [astro-ph]].
  %%CITATION = JCAPA,0901,011;%%


\bbibitem{GKP}
S.~B.~Giddings, S.~Kachru and J.~Polchinski,
``Hierarchies from fluxes in string compactifications,''
Phys.\ Rev.\  D {\bf 66}, 106006 (2002)
[arXiv:hep-th/0105097].
%%CITATION = PHRVA,D66,106006;%%
\bbibitem{DouglasDenef}
F.~Denef, M.~R.~Douglas, B.~Florea, A.~Grassi and S.~Kachru,
``Fixing all moduli in a simple F-theory compactification,''
Adv.\ Theor.\ Math.\ Phys.\  {\bf 9}, 861 (2005)
[arXiv:hep-th/0503124].
%%CITATION = 00203,9,861;%%

\bbibitem{Allenach}
J.~P.~Conlon, C.~H.~Kom, K.~Suruliz, B.~C.~Allanach and F.~Quevedo,
``Sparticle Spectra and LHC Signatures for Large Volume String
Compactifications,''
arXiv:0704.3403 [hep-ph].
%%CITATION = ARXIV:0704.3403;%%  


\bbibitem{bergetal}
M.~Berg, M.~Haack and B.~Kors,
``String loop corrections to Kaehler potentials in orientifolds,''
JHEP {\bf 0511}, 030 (2005)
[arXiv:hep-th/0508043].
%%CITATION = JHEPA,0511,030;%%

    %\cite{Burgess:2003ic}
\bibitem{BKQ}
C.~P.~Burgess, R.~Kallosh and F.~Quevedo,
``de Sitter string vacua from supersymmetric D-terms,''
JHEP {\bf 0310}, 056 (2003)
[arXiv:hep-th/0309187].
%%CITATION = JHEPA,0310,056;%%

%\cite{Saltman:2004sn}
\bibitem{Saltman:2004sn}
A.~Saltman and E.~Silverstein,
``The scaling of the no-scale potential and de Sitter model building,''
JHEP {\bf 0411}, 066 (2004)
[arXiv:hep-th/0402135].
%%CITATION = JHEPA,0411,066;%%

\bbibitem{roulette}
J.~R.~Bond, L.~Kofman, S.~Prokushkin and P.~M.~Vaudrevange,
``Roulette inflation with Kaehler moduli and their axions,''
Phys.\ Rev.\  D {\bf 75}, 123511 (2007)
[arXiv:hep-th/0612197].
%%CITATION = PHRVA,D75,123511;%%



\bbibitem{ourprevious}
J.~Simon, R.~Jimenez, L.~Verde, P.~Berglund and V.~Balasubramanian,
``Using cosmology to constrain the topology of hidden dimensions,''
arXiv:astro-ph/0605371.
%%CITATION = ASTRO-PH/0605371;%%

%\cite{Balasubramanian:2007nu}
%\bibitem{Balasubramanian:2007nu}
V.~Balasubramanian, P.~Berglund, R.~Jimenez, J.~Simon and L.~Verde,
``Topology from Cosmology,''
JHEP {\bf 0806}, 025 (2008)
[arXiv:0712.1815 [hep-th]].
%%CITATION = JHEPA,0806,025;%%


\bbibitem{jimmy}
R.~Holman and J.~A.~Hutasoit,
``Systematics of moduli stabilization, inflationary dynamics and power
spectrum,''
JHEP {\bf 0608}, 053 (2006)
[arXiv:hep-th/0606089].
%%CITATION = JHEPA,0608,053;%%

  R.~Holman and J.~A.~Hutasoit,
``Axionic inflation from large volume flux compactifications,''
arXiv:hep-th/0603246.
%%CITATION = HEP-TH/0603246;%%


\bibitem{08092982}
  A.~C.~Vincent and J.~M.~Cline,
  ``Curvature Spectra and Nongaussianities in the Roulette Inflation Model,''
  JHEP {\bf 0810}, 093 (2008)
  [arXiv:0809.2982 [astro-ph]].
  %%CITATION = JHEPA,0810,093;%%


\bibitem{arXiv:astro-ph/0703486}
  C.~Ringeval,
  ``The exact numerical treatment of inflationary models,''
  Lect.\ Notes Phys.\  {\bf 738}, 243 (2008)
  [arXiv:astro-ph/0703486].
  %%CITATION = LNPHA,738,243;%%


\bibitem{08123500}
  C.~Hikage, K.~Koyama, T.~Matsubara, T.~Takahashi and M.~Yamaguchi,
  ``Limits on Isocurvature Perturbations from Non-Gaussianity in WMAP
  Temperature Anisotropy,''
  Mon.\ Not.\ Roy.\ Astron.\ Soc.\  {\bf 398}, 2188 (2009)
  [arXiv:0812.3500 [astro-ph]].
  %%CITATION = MNRAA,398,2188;%%

\end{thebibliography}
\end{document}